\documentclass[aps,prl,twocolumn,showpacs,amsmath,amssymb]{revtex4}
\usepackage{graphicx}
\usepackage{bm}
\def\Im {\mbox{Im}}
\def\be{\begin{equation}}       \def\ee{\end{equation}}
\def\bea{\begin{eqnarray}}      \def\eea{\end{eqnarray}}

\begin{document}
 \title{An Effective Model  of Magnetoelectricity in Multiferroics $RMn_2O_5$ }
\author{Chen Fang}
\affiliation{Department of Physics, Purdue University, West
Lafayette, Indiana 47907}
\author{Jiangping Hu}
\affiliation{Department of Physics, Purdue University, West
Lafayette, Indiana 47907}
\date{\today}

\begin{abstract}
An effective model is developed to explain the phase diagram and the
mechanism of magnetoelectric coupling in multiferroics, $RMn_2O_5$.
We show that the nature of magnetoelectric coupling in $RMn_2O_5$ is
a coupling between two Ising-type orders, namely, the ferroelectric
order in the b axis, and the    coupled magnetic order between two
frustrated antiferromagnetic chains. The frustrated magnetic
structure drives the system to a commensurate-incommensurate phase
transition, which can be understood as a competition between a
collinear or col-plane order stemming from the `order by disorder'
mechanism and a chiral symmetry order. The low energy excitation is
calculated and the effect of the external magnetic field is
analyzed. Distinct features in the electromagnon spectrums in the
incommensurate phase are predicted.
\end{abstract}


\maketitle

%%%%%%%%%%%%%%%%%%%%%%%%%%%%%%%%%%%%%%%%%%%%%%%%%%%%%%%%%%%%%%%%%%%%%%%%%%%%%%%%%%%
%%%%%%%%%%%%%%%%%%%%%%%%%%%%%%%%%%%%%%%%%%%%%%%%%%%%%%%%%%%%%%%%%%%%%%%%%%%%%%%%%%%%
%%%%%%%%%%%%%%%%%%%%%%%%%%%%%%%%%%%%%%%%%%%%%%%%%%%%%%%%%%%%%%%%%%%%%%%%%%%%%%%%%%%%%
Recently, the search for new spin-electronics materials has led to a
discovery of novel gigantic magnetoelectric and magnetocapacitive
effects in  rare-earth manganites, magnetoelectric multiferroics
\cite{Tokura2006,Cheong2007}. Unlike the magnetic ferroelectroics
studied in 1960s and 1970s where magnetism and ferroelectricity
couple weakly, the magnetism and ferroelectricity in the new
materials couple so strongly that the ferroelectricity  can be
easily manipulated by applying a magnetic field and the magnetic
phase can be controlled by applying an electric field
\cite{Kimura2003a,Hur2004a}. This ease of manipulation promises
great potential for important technological applications in novel
spintronics devices.

The physics of the multiferroics involves the interplay between many
degrees of freedom, such as charge, spin, orbital and lattice.
Tremendous effort has been devoted to decode the fundamental
mechanism of the strong coupling between the magnetism and
ferroelectricity. Experimentally, two major classes of magnesium
oxide multiferroics, have been discovered. The first class  is the
orthorhombic rare-earth manganites $RMnO_3(R = Gd, T b, Dy,
...)$\cite{Kimura2003,Goto2004}, characterized by spiral magnetism
strongly coupled with the ferroelectricity. An effective
Ginzburg-Landau theory incorporating the space group symmetry and
time reversal symmetry has been constructed to explain the
fundamental physics\cite{Mostovoy2006}. Microscopically,
Dzyaloshinskii-Moriya spin-orbit interaction is the underlying
mechanism of the ferroelectricity
\cite{Katsura2005,Katsura2007,Sergienko2006}.  The second class of
materials are the manganese oxides with general formula $RMn_2O_5
(R=Y, Tb, Dy ,...)$
\cite{Kigomiya2003,Hur2004b,Chapon2004a,Kadomtseva2006}. These
insulating materials consist of linked $Mn^{4+}O_6$ octahedra and
$Mn^{3+}O_5$ pyramids with a $Pbam$ space group symmetry. %Due to the
%complexity of the lattice structure in one unit cell,  the magnetic
%structure in $RMn_2O_5$ is much more complicated than that in
%$RMnO_3$.
Unlike that in $RMnO_3$, the ferroelectricity in
$RMn_2O_5$ exists in a collinear or col-plane  magnetic phase,
suggesting that a different mechanism is involved in the interaction
between the ferroelectricity and magnetism.

In this Letter, we develop an effective model  to explain the phase
diagram and the mechanism of magnetoelectric coupling  in $RMn_2O_5
$. Building upon experimental facts and the space group
symmetry\cite{Kigomiya2003,Hur2004a,Hur2004b,Chapon2004a,
Chapon2004,Chapon2006,Kadomtseva2006,Blake2005,Radulov2007,Sushkov2007},
we show that the magnetoelectric interaction is between two Ising
type orders, the ferroelectric order in the b axis and the  coupled
magnetic order between two frustrated antiferromagnetic chains. The
effective model of the magnetism can be derived from a microscopic
model with  nearest-neighbor magnetic exchange. We show that the
effective model nicely captures the phase diagrams of $RMn_2O_5$. At
high temperature, the commensurate (CM) collinear or   col-plane
order
  is stable due to the `order by disorder' mechanism
\cite{Shender1982,Chandra1990,Henley1989} and the existence of an
easy axis. As the temperature decreases,   a chiral symmetry order
replaces the collinear or col-plane  order, and the magnetic
structure becomes incommensurate (ICM). This model predicts that an
external magnetic field along the b-axis can  drive the system from
the ICM phase to the CM phase while that along the a-axis can drive
the system from the CM phase to the ICM phase. The model predicts
the following distinguished properties of low energy excitations:
(1) the emergence of electromagnons in the ICM phase when there is
no electromagnons at the lowest energy in the CM phase; (2) the
presence of distinguished kinks  in the energy dispersions of the
electromagnons, unlike the dispersion of normal phason  in
conventional ICM phase which has a finite energy jump at the half of
ICM wavevector \cite{Bak1982}; (3) a double peak structure at low
energy in optical conductivity due to the absorption of the
electromagnons with a selection rule of the electric field along the
b axis; (4) no new peak splitting  in the electromagnon spectrums in
the presence of external magnetic field along the a or b axis, which
differs from the conventional picture of the Zeeman energy splitting
of magnons; (5) the increase(decrease) of the energy gaps of the
electromagnons as the field increases along the b(a) axis.

{\it  Magnetoelectric coupling:} The ferroelectricity in $RMn_2O_5$
  is substantially different from that in $RMnO_3$. Experiments have
  shown that the ferroelectricity only exists along the b axis in $RMn_2O_5$,
   but can be observed in both the a and c directions in $RMnO_3$. Most importantly,
    recent measurements of optical conductivity have revealed
    opposite selection rules for these two materials. In $RMnO_3$,
    low energy absorption is observed when the electric field is
    perpendicular to the static ferroelectric polarization
    direction\cite{Pimenov2006,Sushkov2007}.
    %Theoretically, this result is consistent with    general mean field
%    theory, since the fluctuation of ferroelectricity around mean field solution along
%    the polarization direction is naturally decoupled from magnetic dynamics up to the second order.
%    The low energy excitations, electromagnons, can only be
%   formed between the magnons and the electric degree of freedom perpendicular to ordered
%   direction.
   The opposite is true in $RMn_2O_5$, namely, low energy absorption is only observed when
   the electric field is polarized along the b axis\cite{Sushkov2007}. This suggests that the electric
   degree of freedoms along the a and c axis simply has no coupling to magnetic degrees
   of freedom at low energy. Only the ferroelectricity along the b axis $P_b$ couples to
   the magnetic degree of freedom. The order parameter $P_b$ is an Ising-type order.
    Therefore, the nature of the magnetoelectric coupling in $RMn_2O_5$ is a coupling
    between two Ising-type orders.( in the following paper,
 we refer the a,b,c axis to the x,y,z axis respectively for conveniences i.e.  $P_y=P_b$).

\begin{figure}
\includegraphics[width=5cm,height=3cm]{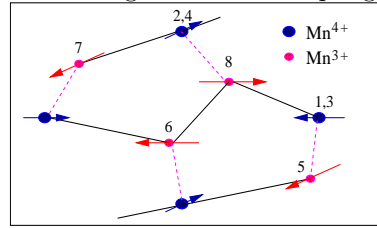}
\caption{\label{twochains}A sketch of spin structures from  a top
view along the c axis in one unit cell of $RMn_2O_5$. The red lines
reflect the frustrated magnetic coupling between two chains. }
\end{figure}

What is the Ising order in the magnetic side? The answer to this
question can be obtained by in-depth analysis of the magnetic
structure and the space group. As shown in
\cite{Blake2005,Chapon2006},the main magnetic structure along the a
axis is two antiferromagnetic chains joined by $Mn^{3+}$ and
$Mn^{4+}$ atoms (see Fig.\ref{twochains}). The antiferromagnetic
coupling between the two chains indicated by the red lines in
Fig.\ref{twochains} are completely frustrated. Therefore, in an
effective model, we at least need two antiferromagnetic orders $\vec
n_1$ and $\vec n_2$ to describe the magnetic physics. A possible
Ising order from these two vector magnetic orders is $\vec
n_1\cdot\vec n_2$. Due to the experiment fact that no magnetic
moment in the c-axis is observed, it naturally leads to the
construction of  the possible lowest order magnetoelectric coupling
bewteen $P_y$ and $\vec n_i,i =1, 2$ as
\begin{eqnarray}\label{coupling}
H_{em}=  \lambda_x P_y n_1^xn_2^x+\lambda_y P_y n_1^yn_2^y
\end{eqnarray}
The possible differences between the coupling parameters $\lambda_x$
and $\lambda_y$ reflects  real lattice structure.

Now we show that Eq.\ref{coupling} is consistent with the space
group analysis.  The space group of $RMn_2O_5$ has been
analyzed\cite{Harris2006}. The lattice of $RMn_2O_5$ belongs to Pbam
structure. With the modulation vector $q =(1/2, 0,k_c)$, the space
group has a single two dimensional irreducible representation in
which the four symmetry lattice transforms can be represented by $I,
m_x = \sigma_x, m_y = \sigma_y,m_xm_y = i\sigma_z$ where $\sigma_i$
are Pauli Matrix. Symmetry adapted variables can be constructed as
linear combinations of spin operators that transform in accordance
with these matrices. The ion spins in one unit cell are numbered one
to eight as shown in Fig.\ref{twochains}. The space inversion
symmetry, together with the experimental facts that $S_1=S_3$,
$S_2=S_4$,$S_5=S_7$, $S_6=-S_8$ and that the spin moments are only
in a-b plane, suggests that the possibilities of magnetoelectric
coupling term can be narrowed down to
\begin{eqnarray} \label{group}&  H_{em}&=  i\lambda_x P_y(-S_{2}^x(q)
S_{6}^{x\star}(q)+S_{1}^x(q) S_{5}^{x\star}(q)-c.c) \nonumber \\& +&
 i\lambda_yP_y(-S_{2}^y(q) S_{6}^{y\star}(q)+S_{1}^y(q)
S_{5}^{y\star}(q)-c.c).
\end{eqnarray}
  Converting
these spin operators to the two antiferromagnetic orders,  we can
simplify Eq.\ref{group} to Eq.\ref{coupling}.

\begin{figure}
\includegraphics[width=5cm]{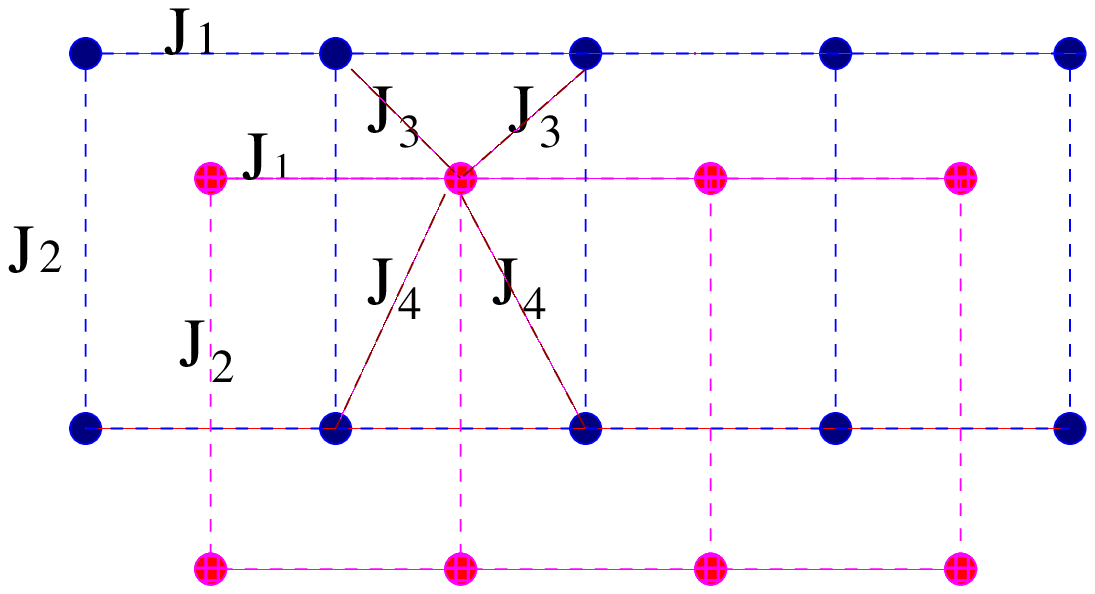}
\caption{\label{lattice}  A sketch of lattice structures of  two
frustrated coupled antiferromagnetic Heisenberg models in two
dimension. }
\end{figure}

{\it Effective magnetic model:}
 In condensed matter physics, an effective model at low energy is largely independent
  of microscopic models if  they  share  same essential physics. Therefore, one can   derive the effective
 model on a much simplified lattice structure. In  the case of $RMn_2O_5$,
the important magnetic physics  along the  a axis are two
antiferromagnetic chains
 with frustrated coupling\cite{Blake2005,Chapon2006}. Experiments have shown  a 1/4 commensurate magnetic wavevector along
 the c axis, which can also be viewed as an antiferromagnetic order if the unit cell
 is doubled along the c-axis. Therefore, we can derive the
 effective model from a microscopic model with two antiferromagnetic
orders defined on the two interpenetrating sublattices as
illustrated in Fig.\ref{lattice}, where  $J_{1,(2)}$  are the
effective antiferromagnetic exchange couplings which establish two
antiferromagnetic orders, $J_3$ is the effective frustrated coupling
between two chains in one unit cell along the a axis and $J_4$ is
the effective frustrated coupling between two chains in two neighbor
unit cells along the c axis. Using standard field
theory\cite{Tsvelik,Henley1989}, we can show that the effective
field theory described by the two antiferromagnetic orders $\vec
n_1$ and $\vec n_2$ is given by the following Hamiltonian
\begin{widetext} \bea \label{magnetic}
H_{m}&&=\int\{\sum_{i}[\frac{\rho_{1s}}{2}(\partial n_{i}/\partial
x)^2+\frac{\rho_{2s}}{2}(\partial n_{i}/\partial
z)^2]+\alpha(n_1\frac{\partial n_2}{\partial x}-n_2\frac{\partial
n_1}{\partial x})-\\\nonumber&&\eta(\frac{\partial n_1}{\partial
x}\frac{\partial n_2}{\partial z}+\frac{\partial n_1}{\partial
z}\frac{\partial n_2}{\partial x})-\tilde{g}(T)(n_1\cdot
n_2)^2-D_0\sum_{i}(n_{i}^x)^2\}dxdz.\eea
\end{widetext}
where $\tilde{g}(T)=\tilde{g}_0+\tilde{g}_1 T$ is a temperature
dependent parameter induced by the quantum and thermal fluctuation,
 the parameter $\alpha$
reflects that the intra frustrated coupling  in one unit cell is
larger than the inter frustrated coupling along the a axis between
two neighbor unit cells along the c axis, and the parameter $D_0$
describes a possible magnetic easy axis along the a axis. From the
microscopic coupling parameter, $
\tilde{g}_1=0.26(J_1+J_2)Sa^{-2}(\frac{J_3+J_4}{J_1+J_2})^2$,
$\tilde{g}_2=2.4\frac{\tilde{g}_1}{(J_1+J_2)S}$, $\rho_{1s}=J_1
S^2$, $\rho_{2s}=J_2 S^2$,
$\alpha=\frac{(J_1+J_2)S^2}{4}\frac{(J_3-J_4)}{(J_1+J_2)a}$ and
$\eta=\frac{(J_1+J_2)S^2}{4}\frac{J_3+J_4}{J_1+J_2}$.
 Adding the
lattice dynamics, we reach the total effective Hamiltonian as
\begin{eqnarray}\label{H2}
H= \int dx dz [( \frac{\kappa}{2}P_y^2+\frac{1}{2M}\pi_y^2)
+H_{{em}}]+H_{{m}}
\end{eqnarray}
where $\pi_y$ is the conjugate momentum of $P_y$.

\begin{figure}
\includegraphics[width=6cm]{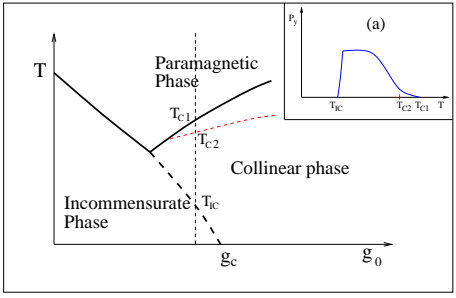}
\caption{\label{phase}   The phase diagram as $g_0$ vs $T$. The
vertical dashed line represents phase transitions in most of
$RMn_2O_5$ materials. The inset (a) is the ferroelectricity as a
function of temperature.}
\end{figure}

The  Hamiltonian in Eq.\ref{H2} precisely captures the phase
diagrams of $RMn_2O_5$\cite{Kigomiya2003,Hur2004a,Chapon2004}.
 To study the magnetic phase diagram,
 we can integrate out the lattice dynamics. After integrating out the lattice dynamics,
 the effective magnetic Hamiltonian is the Hamiltonian in
 Eq.\ref{magnetic} with a replacement of $\tilde g$ by
 $g(T)=g_0+\tilde{g}_1 T$ where
$g_0=\tilde{g}_0+\frac{\lambda_+^2}{8\kappa}$, $D_0$ by
$D=D_0+\frac{\lambda_+\lambda_-}{4\kappa}$ and  an additional term
$-\gamma(n_1^xn_2^x-n_1^yn_2^y)^2$ where $\gamma=
\frac{\lambda_-^2}{8\kappa}$ and
$\lambda_\pm=\lambda_x\pm\lambda_y$. It is clear that the $\alpha$
term favors an ICM phase while  $g(T), \gamma$ and $D$ favor a CM
phase. If $D\neq0$, at relatively high transition temperature
$T_{c1}$ , the model   exhibits a first phase transition to a
collinear magnetic phase with order $<n_1^xn_2^x>\neq 0$ and then
exhibits a second phase transition at $T_{c2}<T_{c1}$ with order
$<n_1^yn_2^y>\neq 0$.  The collinear phase becomes a col-plane
phase. The ferroelectricity is given by
$<P_y>=-\frac{\lambda_x<n_1^xn_2^x>+\lambda_y<n_1^yn_2^y>}{\kappa}$.
With the proper values of $\alpha$, at a low temperature $T_{IC}$,
the ICM phase can win over the col-plane magnetic phase. In the ICM
phase, the global average, $<P_y> =0$. Fig.\ref{phase}  sketches the
phase diagram. The phase diagram qualitatively matches the current
experimental results on the phase diagram of
$RMn_2O_5$\cite{Kigomiya2003,Hur2004a,Chapon2004}. Results from mean
field or large N limit calculation, using the real experimental data
as input, will be reported elsewhere\cite{Hu}. In this paper, we
present a thorough study of the simplified version of the model in
one dimension which captures the essential low energy physics.

{\it Coupled Sine-Gordon model:}   We focus on an one-dimensional
effective model by ignoring the dynamics in z direction in
Eq.\ref{H2}. The one dimensional model still captures essential
physics of the frustration.  For the one dimensional model, we can
go beyond mean field calculation and evaluate the dynamics in a
controllable perturbation manner. Using the angle to parameterize
the antiferromagnetic vector in x-y plane as $\vec
n_i=(cos\theta_i,sin\theta_i)$,  we obtain the effective Hamiltonian
in one dimension  as a coupled Sine-Gordon model, \bea \label{oned}
H& =&\frac{\rho}{2}(\frac{\partial\theta_-}{\partial
x})^2+\alpha\sin\theta_-\frac{\partial\theta_{+}}{\partial
x}-g\cos^2\theta_--\gamma\cos^2\theta_{+} \nonumber
\\
&+&\frac{\rho}{2}(\frac{\partial\theta_+}{\partial
x})^2-D\cos\theta_+\cos\theta_{-},\eea where
$\theta_\pm=\theta_1\pm\theta_2.$ Minimizing the potential part in
Eq.\ref{oned}, the CM-ICM phase transition line is estimated to be
  \bea \label{line}
\rho(g(T)+\gamma/2+D)=\alpha^2/2+\frac{D^2}{2(\alpha^2-\rho
g(T))}.\eea

 To obtain a more accurate result, we can apply a controllable numerical method\cite{Mcmilan1975}.
  In the CM phase, the ground state of the model  is
always described by $\theta_+=\theta_-=0$ or $\pi$. In the ICM
phase, it is clear that the spins on the two antiferromagnetic
chains are both rotating with the same wave vector q throughout the
system. Therefore, we can take the following ansatz: $
\theta_-(x)=\sum_{n=0}^{\infty}(a_n\cos nqx+b_n\sin nqx)$ and $
\theta_+(x)=qx+\sum_{n=0}^{\infty}(c_n\cos nqx+d_n\sin nqx).$
  The coefficient in this ansatz decreases rapidly as $n$ increases.
   As an example, for a typical set of parameters
$\{\rho=2,\alpha=0.4,g=0.03,\gamma=0.01,d=0.005\}$, a numerical
solution of the ICM ground state   is given by
$\theta_+=qx+0.0322\sin(2qx)-0.000255\sin(4qx)+...$ and $
 \theta_-=-\frac{\pi}{2}+0.0526\cos(qx)+0.0000676\cos(3qx)+...,$
where $q=0.199$.  Therefore, the first two terms in  the ansatz can
give us a very good
 approximation of the ground state.

{\it Energy dispersion of magnons in the CM phase:} In the CM phase,
the ferroelectricity, $<P_y>= -\frac{\lambda_++\lambda_-}{2\kappa}$.
The dynamics of magnons and the dynamics of electric degree of
freedoms are decoupled.   To show this, one can expand the free
energy at the vicinity of  the ground state, $ u_1=\delta
\theta_+,\;u_2=\delta\theta_-,\beta=\delta P_y$ and show that there
is no second order coupling between $u_i$ and $\beta$ (the lowest
order coupling between $u_i$ and $\beta$ is $\lambda_+\beta u_2^2+
\lambda_-\beta u_1^2$). Therefore, there is no electromagnon
excitation.  The energy dispersions of the two magnons   in the CM
phase are given by \bea E_{cm}^{\pm}=\sqrt{\rho(\rho
k^2+\gamma+g+D\pm\sqrt{ (g-\gamma)^2 + \alpha^2k^2 })}.\eea

{\it Electromagnons in the ICM phase:} In the ICM phase,
$<P_y(x)>=-\frac{\lambda_+}{2\kappa}cos (qx)$ where $q$ is the ICM
wavector. The magnon and phonon dynamics are coupled due to the ICM
modulation. Therefore, the low energy excitations are
electromagnons. To calculate the dispersion, we expand the
Hamiltonian at the vicinity of the ICM phase, $ u_1=\delta
\theta_+,u_2=\delta\theta_-  $ and $\beta= \delta P_y $. The
fluctuation up to the second order of these dynamic variables is
given by $ \delta H^{(2)}=\delta H^{(2)}_0+\delta H^{(2)}_1$. The
first term is the free part given by\bea \delta
H^{(2)}_0=\sum_{k}(\sum_{i=1}^2
E_i(k)\mu_i(k)\mu_i(-k)+\omega_0^2\beta(k)\beta(-k)) \eea with
$E_1(k)=\rho
k^2+\frac{Da_1}{2}+\frac{\lambda_-(a_1\lambda+\lambda_-)}{4\kappa}$,
$E_2(k)=E_1(k)+\Delta_0 $ where $\Delta_0= \alpha q-2\tilde{g}
-\frac{\lambda_-^2}{4\kappa}$ and the phonon frequency
$\omega_0=\sqrt{\kappa}$ (for convenience, we have taken the mass
$M=1$ in Eq.\ref{H2}). The second term is the interaction part \bea
& &\label{couple} \delta H^{(2)}_1= \int dx [\alpha
a_1\cos(qx)u_1'u_2+D\sin(qx)u_1u_2\nonumber \\
& &+\frac{1}{2}\beta(-\lambda_-u_1sin(qx)+\lambda_+u_2)].\eea

\begin{figure}
\includegraphics[width=7cm]{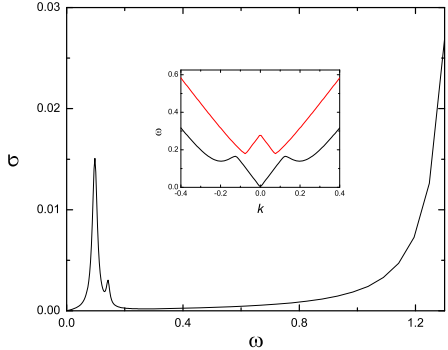}
\caption{\label{absorption}  The numerical result of the optical
conductivity versus frequency  with the parameters set to:
$\{\rho,\alpha,\lambda_+,\lambda_-,\kappa,D,g\}=\{2,0.4,0.06,0.06,2,0.002,0.005\}$.
The inset is the dispersion of the electromagons. }
\end{figure}

In Eq.\ref{couple}, the terms in the first line  couple $u_1(k)$
with $u_2(k\pm q)$ and vice versa and modify the gap $\Delta_0$
between two magnon modes. In general, as shown in
Fig.\ref{absorption}, the coupling creates a distinguished kink in
dispersion curve  around $k=q/2$ rather than a finite energy jump at
$k=q/2$ which is normally expected in the ICM phase described in a
simple Sine-Gordon model\cite{Bak1982}. The terms in the second line
describe the coupling between the magnon and the phonon.  The
coupling results in two general effects. First, the coupling leads
to a change of phonon frequency. Under condition $\omega^2_0>>\rho
E_i(0)$, the shift frequency of the phonon is roughly given by
$\delta\omega^2=\frac{\rho\lambda_+^2}{4(\omega^2_0-\rho
E_2(0))}+\frac{\rho\lambda_-^2}{8(\omega^2_0-\rho E_2(q))}$. Second,
the coupling allows us to measure  the magnons in optical
conductivity. The optical conductivity  is given by $
\sigma(\omega)=\omega\Im[G_{\beta\beta}(\omega,0)],$ where
$G_{\beta\beta}(\omega,k)$ is the full propagator of the $\beta$. In
our model, we expect  double peaks in the optical conductivity at
the gap energy of the two magnons in the ICM phase. Up to the second
order,   we have,
$\Im[G_{\beta\beta}(\omega,0)]=\pi[\rho\frac{\lambda_-^2}{8\omega_0^4}\delta(\omega^2-\rho
E_1(q))+ \rho\frac{\lambda_+^2}{4\omega_0^4}\delta(\omega^2-\rho
E_2(0))+(1-\rho\frac{\lambda_-^2/2+\lambda_+^2}{4\omega_0^4})\delta(\omega^2-\omega_0^2-\delta\omega^2)]$
.   In Fig.\ref{absorption}, by numerically solving the dynamical
equations of $\delta H^{(2)}$, we plot the  result of
$\sigma(\omega)$ and the dispersions of electromagnons with a
typical parameter setting
$\{\rho,\alpha,\lambda_+,\lambda_-,\kappa,D,g\}=\{2,0.4,0.06,0.06,2,0.002,0.005\}$.

 %n Fig.\ref{absorptin}, we plot the optical
%conductivity in the ICM phase for certain parameters. The term
%couples $u_1(k)$ with $u_2(k\pm q)$ and vice versa, which further
%increase the gap between two modes. However, unlike that in the ICM
%phase described by single Sine-Gordon model\cite{Bak1982}, the $V$
%term does not create a finite energy jump at $q$.
%
%\bea \delta H^{(2)}=\sum_{k,i=1}^2 \epsilon_i(k)u_i(k)u_i(-k)+V \eea
%where $V= \int dx (\alpha a_1\cos(qx)u_1'u_2+D\sin(qx)u_1u_2)$,
%$\epsilon_1(k)=\frac{\rho}{2} k^2+\gamma+Da_1/4$ and
%$\epsilon_2(k)=\epsilon_1(k)+\Delta$ with
%$\Delta_0=q\alpha/2-\gamma-g>0$. The $V$ term couples $u_1(k)$ with
%$u_2(k\pm q)$ and vice versa, which further increase the gap between
%two modes. However, unlike that in the ICM phase described by single
%Sine-Gordon model\cite{Bak1982}, the $V$ term does not create a
%finite energy jump at $q$.

{\it Effects of external magnetic field:} In our model, since $\vec
n$ is the staggered moment field, the effect of an external magnetic
field $H$ is equivalent to creating an easy plane. However, since
 the staggered moment is in the a-b plane, the effect of the
field is only important when the field is along the a or b axis. In
the presence of the easy axis parameter $D$ along the a axis, the
effects of the external magnetic fields along the a axis,$H_a$, and
b axis, $H_b$,  have exact opposite effects.  They simply change $D$
to \bea \label{magnetic}D(H_a, H_b)=D- H_a^2/2\rho+H_b^2/2\rho \eea
in Eq.\ref{oned}.
%The
%effect of the electric field, $\varepsilon$, along the b axis is
%very clear in our model as well. it breaks the degeneracy of two
%ground states of the Ising magnetic order i.e. $\theta_+=\theta_-=0$
%vs $\theta_+=\theta_-=\pi$ since it induces a term $  \epsilon
%\varepsilon  (\lambda_a n_1^an_2^a+\lambda_b n_1^bn_2^b)$ where
%$\epsilon$ is the dielectric constant.
 From  $E_1(k)$,
$E_2(k)$, Eq.\ref{couple} and Eq.\ref{line}, Eq.\ref{magnetic} leads
to a few important and immediate predictions. First, the effects of
$H_a$ and $H_b$ do not depend on their  directions along their own
axis. Second,     $H_b$  can drive the system from the ICM phase to
the CM phase while $H_a$  can drive the system from the CM phase to
the ICM phase. This result has been observed experimentally in
\cite{Hur2004b}. The critical fields to drive the transition can be
estimated for the one dimensional system from Eq.\ref{line}. Third,
in the ICM phase, the energy dispersions of the electromagnons as a
function of $H_a$ and $H_b$ can be predicted. From $E_1(k)$,
$E_2(k)$ and Eq.\ref{couple}, the energy of the electromagnons is
expected to increase (decrease) as $H_b$ ($H_a$) increases. Finally,
the external magnetic field does not add additional peaks, which
contradicts the conventional picture of the Zeeman energy
splitting of magnons. %The physical reason is that there  is no
%degeneracy on the electromagnon modes due to large anisotropy in the
%c axis.

In conclusion,  we develop an effective   model that explains the
phase diagram and the mechanism of magnetoelectric coupling in
multiferroics $RMn_2O_5$. To our knowledge, this is the first
theoretical effective model for these materials.
 A detailed
study of low energy excitations in one dimension  is performed to
explain the selection rules of electromagnon in optical conductivity
measurements\cite{Sushkov2007}. Our prediction of the electromagnon
dispersion  and its dependence on the external magnetic field can be
tested in future experiments. A quantitative study of our model
incorporating  the experimental data  on different $RMn_2O_5$
materials will be reported elsewhere\cite{Hu}. We  expect that the
model presented here can be applied to other multiferronics
materials where ferroelectricity is correlated to a collinear or
col-plane magnetic phase.

{\it Acknowledgments} J. Hu  thanks  S.~Kivelson and S. Brown  for
teaching some concepts presented in this paper. We thank A. B.
Sushkov, R.V. Aguilar and D. Drew for important comments and
extremely useful discussion. We also thank M. Mostovoy for useful
discussion in APS March meeting (2007). This work was supported by
the National Science Foundation under grant number PHY-0603759.

\bibliography{ferroics}

\end{document}